\documentclass[a4paper]{jpconf}

\usepackage{graphicx}
\usepackage{mathrsfs}
\usepackage[intlimits,centertags]{amsmath}
\usepackage{amssymb,amsfonts}
\usepackage[pdftex]{hyperref}
\usepackage[x11names]{xcolor}
\usepackage{aas_macros}
\usepackage[a4paper, left=2.5cm,right=2.5cm,top=3cm,bottom=3cm]{geometry}

\hypersetup{pdftitle={The Dipole Anisotropy of Galactic Cosmic Rays},
pdfsubject={The Dipole Anisotropy of Galactic Cosmic Rays},
pdfauthor={Markus Ahlers},
pdfstartview={FitH},
colorlinks=true,
bookmarksopen=false,
bookmarksnumbered=false,
bookmarksopenlevel=0,
linkcolor=Blue1!60!black,
citecolor=Green1!50!black,
urlcolor=Blue1!70!black
}

\begin{document}
\title{The Dipole Anisotropy of Galactic Cosmic Rays}

\author{Markus Ahlers}

\address{Niels Bohr International Academy \& Discovery Centre, Niels Bohr Institute,\\University of Copenhagen, Blegdamsvej 17, DK-2100 Copenhagen, Denmark}

\ead{markus.ahlers@nbi.ku.dk}

\begin{abstract}
The arrival directions of Galactic cosmic rays exhibit anisotropies up to the level of one per-mille over various angular scales. Recent observations of TeV--PeV cosmic rays show that the dipole anisotropy has a strong energy dependence with a phase-flip around 100~TeV. We argue that this behavior can be well understood by the combination of various effects: the anisotropic diffusion of cosmic rays, the presence of nearby sources, the Compton-Getting effect from our relative motion and the reconstruction bias of ground-based observatories. 
\end{abstract}

\section{Introduction}
\medskip

The arrival directions of Galactic cosmic rays (CRs) are highly isotropic. This is expected from the presence of turbulent magnetic fields in our Galaxy with a root-mean-square field strength at the level of micro-Gauss. For CR nuclei in the TeV--PeV energy range and charge $Z$ the maximal gyroradius can be estimated as $r_g \simeq 1.1\,Z\,{\mathcal{R}_{\rm PV}}\,{B^{-1}_{\mu\rm G}}\,{\rm pc}$, where $\mathcal{R}_{\rm PV}$ is the rigidity in units of peta-Volts and $B_{\mu\rm G}$ the magnetic field strength in units of micro-Gauss. This length scale is much smaller than the typical distance of CR sources, presumably supernova remnants (SNRs)~\cite{Baade1934}, and a CR nucleus is therefore expected to encounter many magnetic scattering centers before arrival at Earth. This diffusive process can also account for the observed steepness of the CR power-law spectrum, in contrast to the hard spectrum expected from diffusive shock acceleration~\cite{Bell1978a,Blandford1978}.

\begin{figure}[t]\centering
\includegraphics[width=0.5\linewidth]{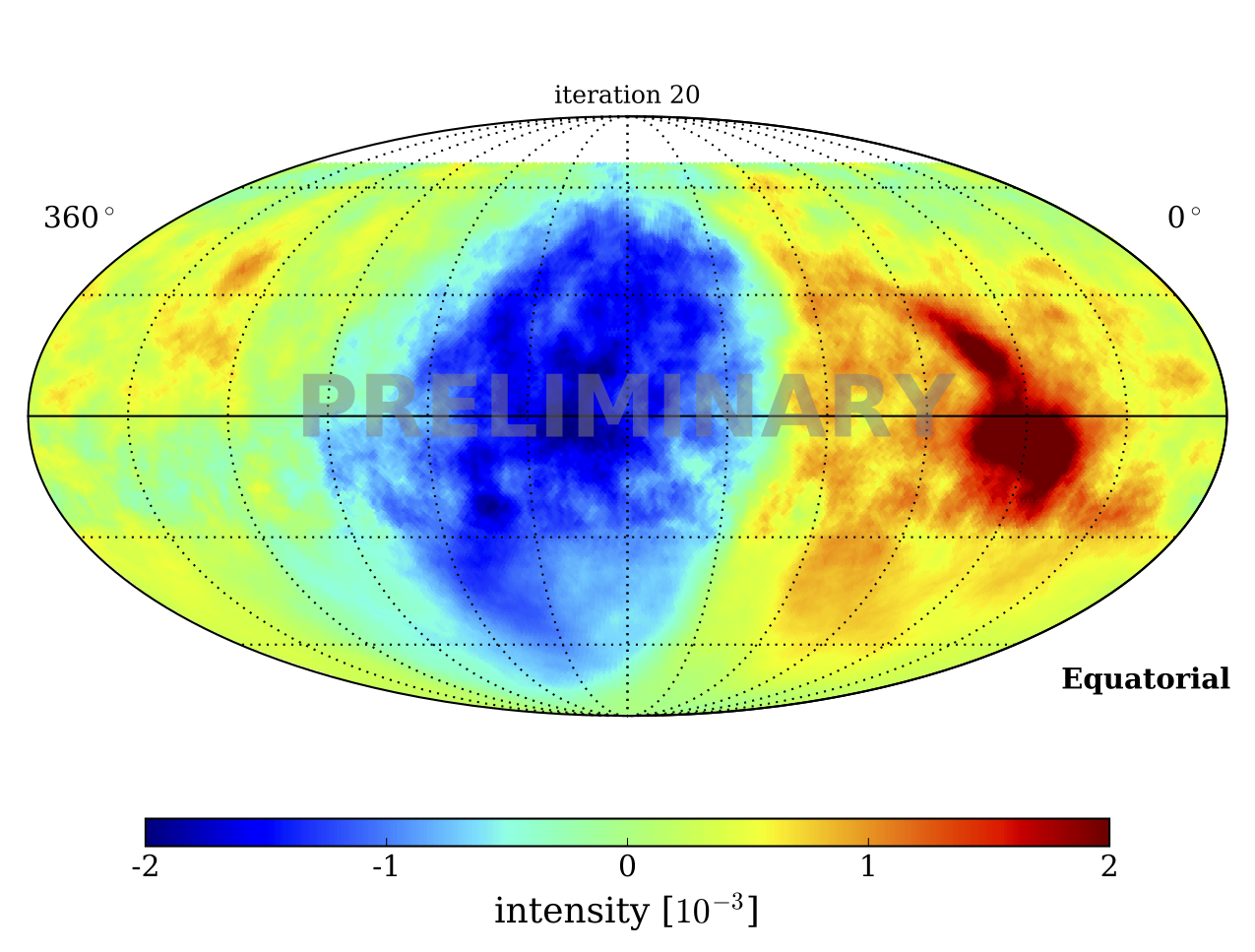}\includegraphics[width=0.5\linewidth]{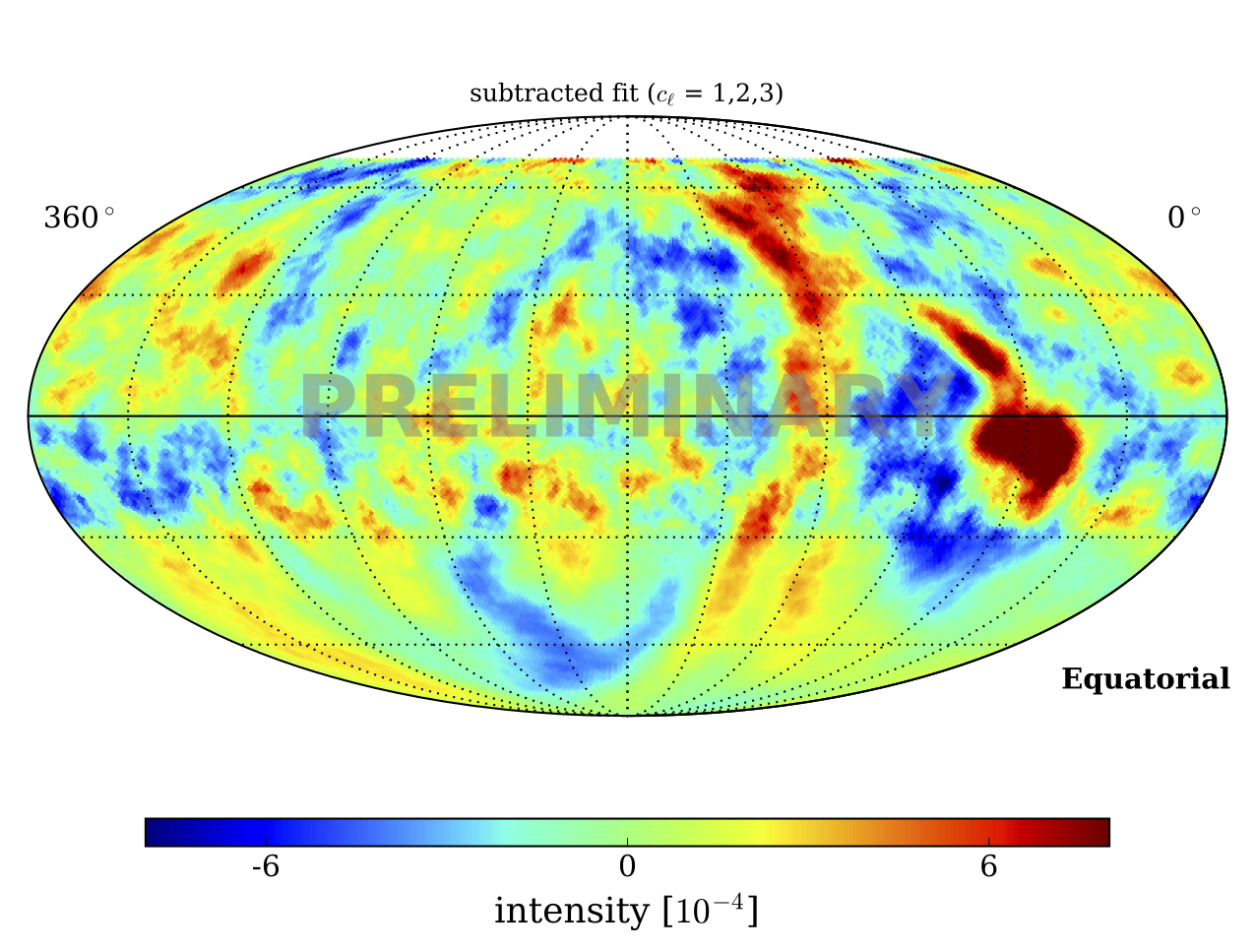}
\caption[]{Preliminary results of the 10~TeV CR anisotropy from a combined analysis of HAWC ($-30^\circ\leq\delta\leq64^\circ$) and IceCube ($-90^\circ\leq\delta\leq-20^\circ$) data. The analysis is based on an iterative maximum-likelihood method introduced in Ref.~\cite{Ahlers:2016njl}. The left map show the full anisotropy in the equatorial coordinate system. The right map shows the same anisotropy after removal of the dipole ($\ell=1$), quadrupole ($\ell=2$) and octopole ($\ell=3$). (Figures from Ref.~\cite{TheHAWC:2017uyf})}\label{fig:IceCubeHAWC}
\end{figure}

However, CR diffusion predicts a weak anisotropy that is proportional to the CR density gradient and the length scale of diffusion. In recent years, many experiments achieved the necessary level of statistics to be able to find anisotropies of one per-mille and below, see {\it e.g.}~Refs.~\cite{DiSciascio:2014jwa,Ahlers:2016rox}. These data revealed that, besides a dominant dipole anisotropy expected from diffusion theory, there is also evidence for medium- and small-scale structures in the maps of CR arrival directions, that extend down to angular scales of 10 degrees. Figure~\ref{fig:IceCubeHAWC} shows a recent (preliminary) anisotropy map from a combined analysis of HAWC and IceCube data that illustrates the complexity of the pattern~\cite{TheHAWC:2017uyf}.

The appearance of medium- and small-scale features is most likely related to non-diffusive CR streaming on length scales much smaller than the effective diffusion length. Various authors have considered the effect of the heliosphere~\cite{Lazarian:2010sq,Desiati:2011xg,Drury:2013uka,Zhang:2014dsu}, the local realization of turbulence~\cite{Battaner:2010bd,Giacinti:2011mz,Ahlers:2013ima,Pohl:2015fdp,Lopez-Barquero:2015qpa,Ahlers:2015dwa} or modifications of CR transport via non-uniform pitch-angle diffusion~\cite{Malkov:2010yq,Giacinti:2016tld}. We refer to the recent review~\cite{Ahlers:2016rox} for further details. 

In this brief review we will focus on the interpretation of the Galactic dipole anisotropy. Diffusion theory predicts that this large-scale anisotropy is proportional to the CR gradient and the effective diffusion length. Na\"ively, one would expect that its orientation is towards the Galactic center and its amplitude is rising as a simple power-law according to the rigidity dependence of the diffusion length. However, this is not supported by recent TeV--PeV CR data. Nevertheless, we will show that the data can be understood in the context of diffusion theory if one accounts for additional effects, namely, the anisotropic diffusion of CRs, the presence of nearby SNRs as the sources of CRs, the Compton-Getting effect from our relative motion in the rest frame of diffusion and a reconstruction bias of ground-based observatories. 

\section{Anisotropy Reconstruction}
\medskip

We assume in the following that the TeV--PeV CR flux is constant over the livetime of observatories. The angular distribution can then be expressed as a function of celestial longitude $\alpha$ (right ascension) and latitude $\delta$ (declination) as, 
\begin{equation}\label{eq:phi}
  \phi(\alpha,\delta) = \phi^{\rm iso}I(\alpha,\delta)\,,
\end{equation} 
where $\phi^{\rm iso}$ corresponds to the isotropic flux level (in units of ${\rm cm}^{-2}\, {\rm s}^{-1}\, {\rm sr}^{-1}$) and $I(\alpha,\delta)$ is the relative intensity of the flux as a function of position in the sky. The anisotropy is defined as the deviation $\delta I = I-1\ll 1$. The reconstruction of Galactic CR anisotropies at the level of one per-mille or less requires the collection of hundreds of millions of events with detectors that are precisely calibrated. It is challenging to satisfy both of these conditions simultaneously. Large event numbers can be achieved by ground-based observatories, that detect CR air showers via surface arrays or underground detectors~\cite{DiSciascio:2014jwa}. Since the atmosphere is part of the detector, a precise determination of the detector response including small variations in the local acceptance is difficult to achieve from simulations alone. However, one can overcome some of these limitations by using the CR data for calibration.

In the local coordinate system of the ground-based observatory the arrival direction of a CR is parametrized by its azimuth angle $\varphi$ (from the north, increasing to the east) and zenith angle $\theta$,
\begin{equation}
{\bf n}'=(\cos\varphi\sin\theta,-\sin\varphi\sin\theta,\cos{\theta})\,.
\end{equation}
The arrival direction in the local horizontal coordinate system is related to the arrival direction in the right-handed equatorial coordinate system via a time-dependent rotation, ${\bf n}={\bf R}(t){\bf n}'$, where the unit vector ${\bf n}$ is given as
\begin{equation}\label{eq:unitn}
{\bf n}=(\cos\alpha\cos\delta,\sin\alpha\cos\delta,\sin{\delta})\,.
\end{equation}
The rotation matrix ${\bf R}$ depends on the local sidereal time $t$ of the observation, which is equivalent to the right ascension angle of the zenith, $\alpha_{\rm zen}(t) = \omega t$, where the sidereal angular frequency $\omega = \omega_{\rm solar} + \omega_{\rm orbit}$ is a combination of the solar angular frequency $\omega_{\rm sol} = 2\pi/24$h and Earth's orbital angular frequency $\omega_{\rm orbit} = 2\pi/1$yr. For an experiment located at a geographic latitude $\Phi$, the transformation is then given by
\begin{equation}\label{eq:Rmatrix}
{\bf R}(t) =
\begin{pmatrix}
  -\cos \omega t\sin \Phi&-\sin \omega t\sin\Phi&\cos\Phi \\
  \sin \omega t&-\cos \omega t&0\\
  \cos \omega t\cos\Phi&\sin \omega t\cos\Phi&\sin\Phi
\end{pmatrix}\,.
\end{equation}
At any time, the ground-based observatory covers an instantaneous field of view which is typically characterized by a maximal zenith angle range, $\theta\leq\theta_{\rm max}$. After many sidereal days of operation, the detector will then observe an integrated field of view characterized by a declination band, $\delta_{\rm min}<\delta<\delta_{\rm max}$, bounded by $\delta_{\rm min} = {\rm max}(-90^\circ,\Phi-\theta_{\rm max})$ and $\delta_{\rm max} = {\rm min}(90^\circ,\Phi+\theta_{\rm max})$.

As indicated earlier, the local detector exposure $\mathcal{E}$ is expected to depend on atmospheric and other time-dependent effects and, in most cases, it is difficult to determine from first principles at a level of one per-mille or less. However, after observation over many sidereal days, the events collected during a fixed sidereal time bin $[t,t+\Delta t]$ are related to a time-integrated detector exposure, where local variations in the detector acceptance can usually be approximated by the average relative acceptance. To be more concrete, the time-integrated local detector exposure $\mathcal{E}$ (accumulated over many sidereal days) is expected to be a product of its angular-integrated exposure $E$ per sidereal time (units of ${\rm cm}^2\, {\rm sr}$) and relative acceptance $\mathcal{A}$ (units of ${\rm sr}^{-1}$),
\begin{equation}\label{eq:E}
  \mathcal{E}(t,\varphi,\theta) \simeq E(t)\mathcal{A}(\varphi,\theta)\,,
\end{equation}
with normalization condition $\int{\rm d}\Omega \mathcal{A}(\Omega)=1$. This ansatz assumes that the relative acceptance of the detector does not strongly depend on sidereal time and time-variations can be compensated by an overall rescaling of the angular-integrated exposure. Note that this is a typical assumption of anisotropy reconstruction methods, like the methods of direct integration~\cite{Atkins:2003ep} or time-scrambling~\cite{Alexandreas1993}. 

The expected number of CRs at a sidereal time $t$ from an azimuth angle $\varphi$ and zenith angle $\theta$ can now be expressed as
\begin{equation}\label{eq:mu}
\mu(t,\varphi,\theta) = \Delta t \Delta\Omega \,I(\alpha(t,\varphi,\theta),\delta(t,\varphi,\theta))  \,\mathcal{N}(t) \,\mathcal{A}(\varphi,\theta)\,,
\end{equation}
where $\mathcal{N}(t) \equiv \phi^{\rm iso}{E}(t)$ gives the expected rate of isotropic background events at sidereal time $t$. The previous relation allows us to simultaneously reconstruct the relative CR intensity $I$, isotropic background rate $\mathcal{N}$ and relative acceptance $\mathcal{A}$ via a fit to the data $n(t,\varphi,\theta)$. For mid-latitude detectors, where the instantaneous field of view is rapidly changing, one can derive the anisotropy via an iterative maximum-likelihood analysis, as shown in Ref.~\cite{Ahlers:2016njl}.

Finally, the reconstructed anisotropy $\delta I$ can be analyzed in terms of its spherical harmonics expansion defined via 
\begin{equation}\label{eq:Ylm}
\delta I(\alpha,\delta) = \sum_{\ell\geq1}\sum_{m=-\ell}^\ell a_{\ell m}Y^{\ell m}(\pi/2-\delta,\alpha)\,,
\end{equation}
where the complex expansion coefficients are related by $a_{\ell\text{-}m} = (-1)^ma^*_{\ell m}$. Unfortunately, due to the observatory's limited field of view ($\delta_{\rm min}\leq\delta\leq\delta_{\rm max}$), this expansion (\ref{eq:Ylm}) can not be inverted unless we introduce an {\it a priori} truncation scale in the multipole expansion. However, there is evidence of small-scale anisotropy in CR data and this truncation seems not to be well motivated. We therefore expect that the multipole reconstruction reported by various observatories in the North and South can show strong variation introduced by the cross-talk of multipoles in the multipole reconstruction.

There is an additional obstacle related to the analysis method of ground-based observatories where the CR data is used for the calibration of the detector. Since this is an important limitation that is often overlooked, we provide in the following section a rigorous mathematical derivation of this effect (see also Appendix of Ref.~\cite{Ahlers:2018qsm}).

\section{Observational Bias}
\medskip

The events recorded at a fixed position $(\varphi, \theta)$ in the local coordinate system can only probe the CR flux along a constant declination $\delta(\varphi, \theta)$, {\it i.e.}, only variations of the flux with respect to right ascension $\alpha(t,\varphi, \theta)$ as the sidereal time increases. Hence, the expectation values (\ref{eq:mu}) are invariant under the simultaneous rescaling
\begin{eqnarray}
\label{eq:scaleI}
  I(\alpha,\delta) \,\,\,\to & I'(\alpha,\delta)  &\equiv\,\,\, I(\alpha,\delta) /a(\delta)/b\,,\\\label{eq:scaleN}
  \mathcal{N}(\varphi,\theta) \,\,\,\to &{\mathcal{N}'(\varphi,\theta)}&\equiv \,\,\,\mathcal{N}(\varphi,\theta)bc
\,,\\\label{eq:scaleA}
  \mathcal{A}(\varphi,\theta) \,\,\,\to &{\mathcal{A}'(\varphi,\theta)}&\equiv\,\,\,{\mathcal{A}(\varphi,\theta)a(\delta(\varphi, \theta))}/{c}\,,\end{eqnarray}
where $a(\delta)$ is an arbitrary function of declination. The normalization factors $b$ and $c$ are defined by the normalization conditions $\int{\rm d}\Omega \mathcal{A}'(\Omega)=1$ and $\int{\rm d}\Omega \delta I'(\Omega)=0$. The symmetry transformations in Eqs.~(\ref{eq:scaleI}), (\ref{eq:scaleN}) and (\ref{eq:scaleA}) imply that the relative intensity $I$ can only be determined up to a declination-dependent function. As we will show in the following, this makes it impossible to infer azimuthally-symmetric ($m=0$) anisotropies, if the CR data are also used for detector calibration. 

We can solve Eq.~(\ref{eq:scaleI}) for the rescaled anisotropy as
\begin{equation}\label{eq:dIp}
\delta I'(\alpha,\delta) = \frac{1+\delta I(\alpha,\delta)}{a(\delta)b}-1\,,
\end{equation}
where the normalization factor $b$ has the explicit form
\begin{equation}
b = \frac{1}{4\pi}\int{\rm d}\Omega\frac{1+\delta I(\alpha,\delta)}{a(\delta)}\,.
\end{equation}
The family of solutions implied by the transformation (\ref{eq:dIp}) also include the true anisotropy, denoted by $\delta \widehat{I}(\alpha,\delta)$ in the following. In order to extract a particular member of the family we have to introduce an additional condition on $\delta I$ that is consistent with the normalization condition, $\int {\rm d}\Omega \delta I(\alpha,\delta) = 0$, but breaks the symmetry~(\ref{eq:scaleI}). The natural choice of this ``fixing'' condition is
\begin{equation}\label{eq:dIfix}
\int{\rm d}\alpha \delta I^{\rm fix}(\alpha,\delta) = 0\,.
\end{equation}
Note that this condition singles out a unique solution. Two solutions, $\delta I'^{\rm fix}$ and $\delta I^{\rm fix}$, imply $a(\delta)b=1$ (cf.~Eq.~(\ref{eq:dIp})) and are therefore identical. On the other hand, if we had found a solution $\delta I$ that does not obey condition (\ref{eq:dIfix}), we can find the fixed solution $\delta I^{\rm fix}$ by the transformation (\ref{eq:dIp}) using the rescaling function
\begin{equation}\label{eq:fix}
a(\delta)b = \frac{1}{2\pi}\int{\rm d}\alpha(1+\delta I(\alpha,\delta))\,.
\end{equation}

Now, the true anisotropy, $\delta \widehat{I}$, can be expanded into spherical harmonics as in Eq.~(\ref{eq:Ylm}) with coefficients $\widehat{a}_{\ell m}$. Applying the fixing condition (\ref{eq:dIfix}) to $\delta \widehat{I}$ implies the transformation (\ref{eq:dIp}) with scaling function
\begin{equation}
a(\delta)b = 1+\sum_{\ell\geq1} \widehat{a}_{\ell 0}\sqrt{\frac{2\ell+1}{4\pi}} P_\ell(\sin\delta)\,.
\end{equation}
But this expression is simply the azimuthally-averaged relative intensity, $\langle\widehat{I}\,\rangle(\delta)$. From Eq.~(\ref{eq:dIp}) we therefore arrive at the (unique) solution
\begin{equation}
\delta I^{\rm fix}(\alpha,\delta) = \frac{1}{\langle\widehat{I}\,\rangle(\delta)}\sum_\ell\sum_{m\neq 0} \widehat{a}_{\ell m} Y^{\ell m}(\pi/2-\delta,\alpha)\,.
\end{equation}
Since the azimuthally-averaged relative intensity is dominated by the monopole, $\langle\widehat{I}\,\rangle(\delta)\simeq 1$, this expression approximates the spherical harmonic expansion of the true anisotropy, $\delta\widehat{I}$, except for azimuthally-symmetric components with $m=0$.

In summary, the anisotropy reconstruction by ground-based observatories, that use the same CR data to determine the local detector acceptance, is insensitive to the $m=0$ components of the spherical harmonics expansion of the anisotropy in the equatorial coordinate system. This observational bias was first pointed out in Ref.~\cite{Iuppa:2013pg} and has important consequences for the interpretation of CR anisotropies.

\section{Dipole Anisotropy}
\medskip

The dipole anisotropy is the leading-order effect predicted from CR diffusion. The dipole term is usually parametrized as 
\begin{equation}\label{eq:dipole1}
\delta I_{\rm dipole}(\alpha,\delta) = \boldsymbol{\delta}\!\cdot\!{\bf n}(\alpha,\delta)\,,
\end{equation}
where ${\bf n}$ is the unit vector introduced in Eq.~(\ref{eq:unitn}). The dipole vector $\boldsymbol{\delta}$ can be expressed in terms of the expansion coefficients of the spherical harmonics (\ref{eq:Ylm}) as
\begin{equation}\label{eq:deltageneral}
\boldsymbol{\delta} \equiv \big(\delta_\text{0h},\delta_\text{6h},\delta_\text{N}\big) =\sqrt{\frac{3}{2\pi}}\big(-\Re (a_{11}),\Im (a_{11}),a_{10}\big)\,.
\end{equation}
Here, we introduced the notation $\delta_\text{0h}$ and $\delta_\text{6h}$ corresponding to the dipole components parallel to the equatorial plane and pointing to the direction of the local hour angle 0h ($\alpha=0^\circ$) and 6h ($\alpha=90^\circ$) of the vernal equinox, respectively. We also introduce $\delta_\text{N}$ as the dipole component pointing to north.

\begin{figure}[t]\centering
\includegraphics[width=0.8\linewidth]{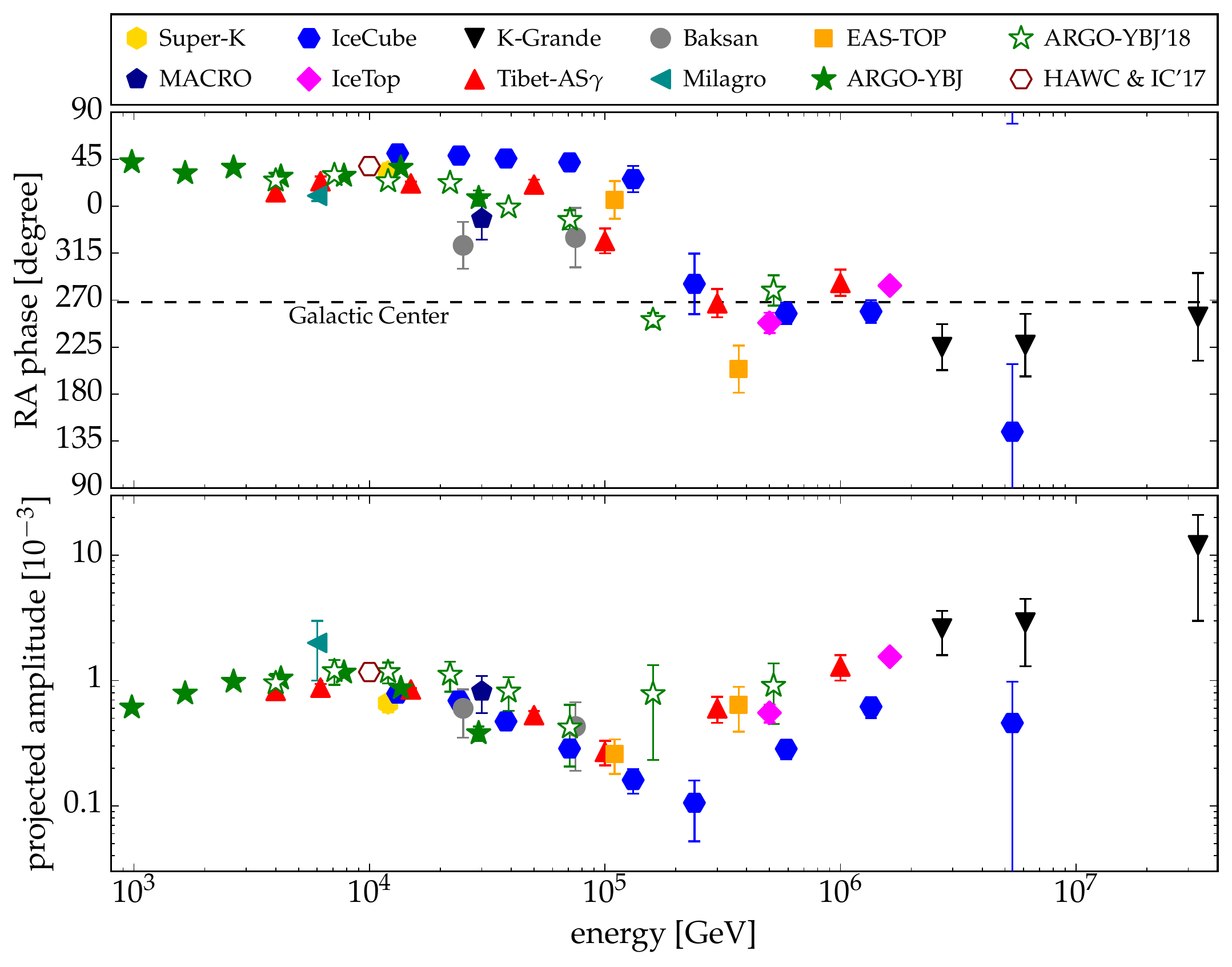}
\caption[]{The phase (top panel) and amplitude (bottom panel) of the projected dipole anisotropy from recent CR measurements~\cite{Ahlers:2016rox,Bartoli:2018ach,TheHAWC:2017uyf}. The dashed line in the top panel shows the right ascension angle of the Galactic center.}\label{fig:dipoledata}
\end{figure}

As discussed in the previous section, ground-based observatories only report the dipole components aligning with the equatorial plane (cf.~Eq.~(\ref{eq:deltageneral})) as
\begin{equation}\label{eq:ampphase}
\big(\delta_\text{0h},\delta_\text{6h}\big) = (A_1\cos\alpha_1, A_1\sin\alpha_1)\,.
\end{equation}
The observed dipole amplitude corresponds to the projection of the true dipole onto the equatorial plane, $A_1= \cos\delta_1 A = \sqrt{3/2\pi}|a_{11}|$. Only the phase $\alpha_1$ corresponds to the true right ascension angle. Figure~\ref{fig:dipoledata} shows a summary of the projected dipole amplitude and dipole phase from various experiments. One can notice a phase-flip and a drop in the amplitude around at an energy of about 100~TeV. We will argue in the following that this behavior can be understood in terms of standard CR diffusion, after accounting for strong anisotropic diffusion in our local environment and the presence of local CR sources.

Fick's law of diffusion theory predicts that the dipole anisotropy is proportional to the spatial gradient of the CR density $\nabla n^\star$ and the diffusion tensor ${\bf K}$,
\begin{equation}\label{eq:diffusedipole}
{\boldsymbol \delta^\star} = \frac{3{\bf K}\!\cdot\!\nabla n^\star}{ n^\star}\,.
\end{equation}
In general, the diffusion tensor is expected to be invariant under rotations along the orientation of the local ordered magnetic field and can be written in the form
\begin{equation}\label{eq:K}
{K}_{ij} = \kappa_\parallel{\widehat{B}_i\widehat{B}_j}+\kappa_\perp({\delta_{ij}-\widehat{B}_i\widehat{B}_j})+\kappa_A{\epsilon_{ijk}\widehat{B}_k}\,.
\end{equation} 
Here, $\hat{\bf B}$ is a unit vector pointing in the direction of the regular magnetic field, $\kappa_\parallel$ and $\kappa_\perp$ denote the diffusion strength along and perpendicular to the magnetic field, respectively, and $\kappa_A$ is the axial diffusion strength (see, {\it e.g.}, Ref.~\cite{Bhatnagar:1954zz}). 

\begin{figure}[t]\centering
\includegraphics[height=0.7\linewidth]{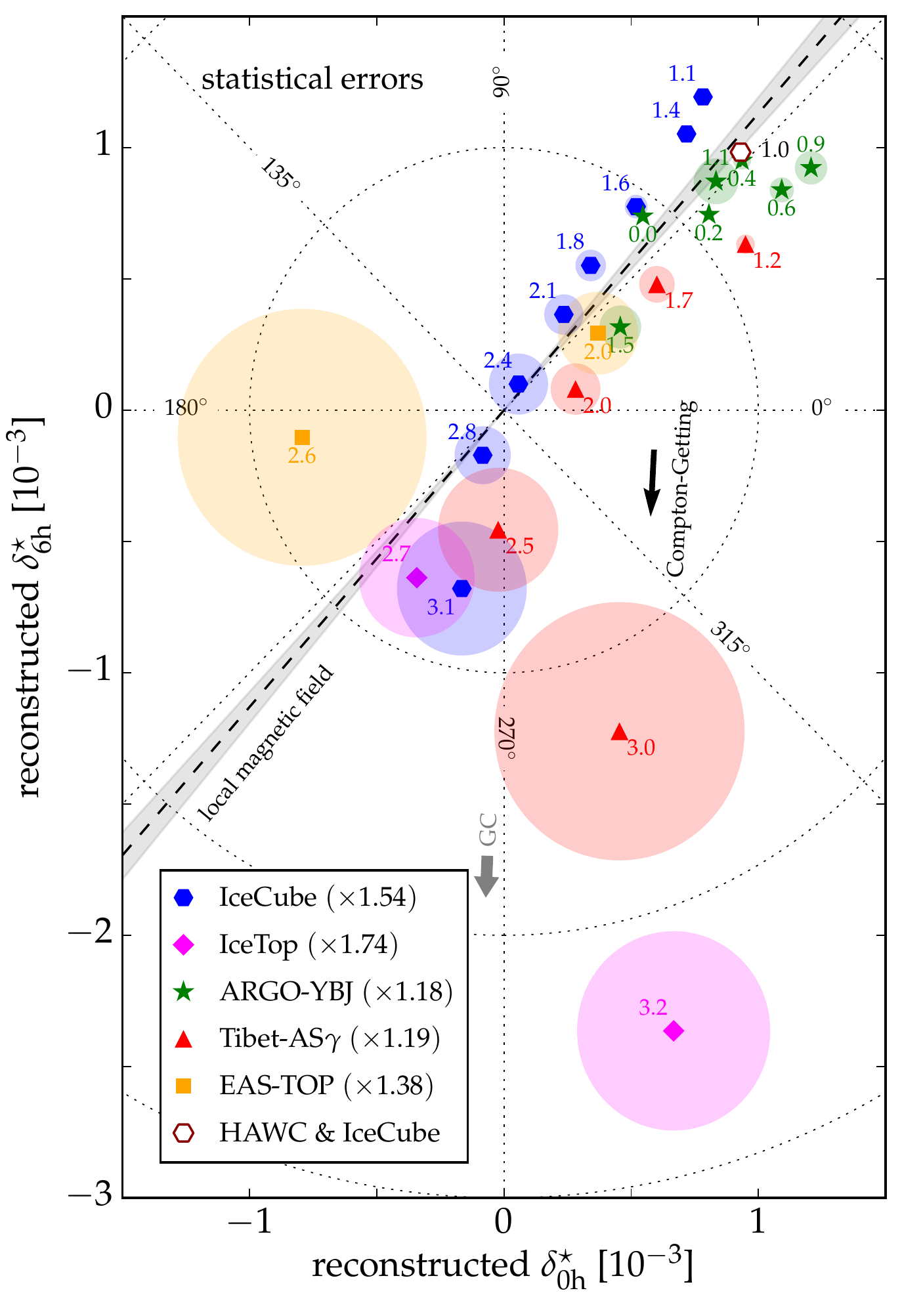}\hfill\includegraphics[height=0.7\linewidth]{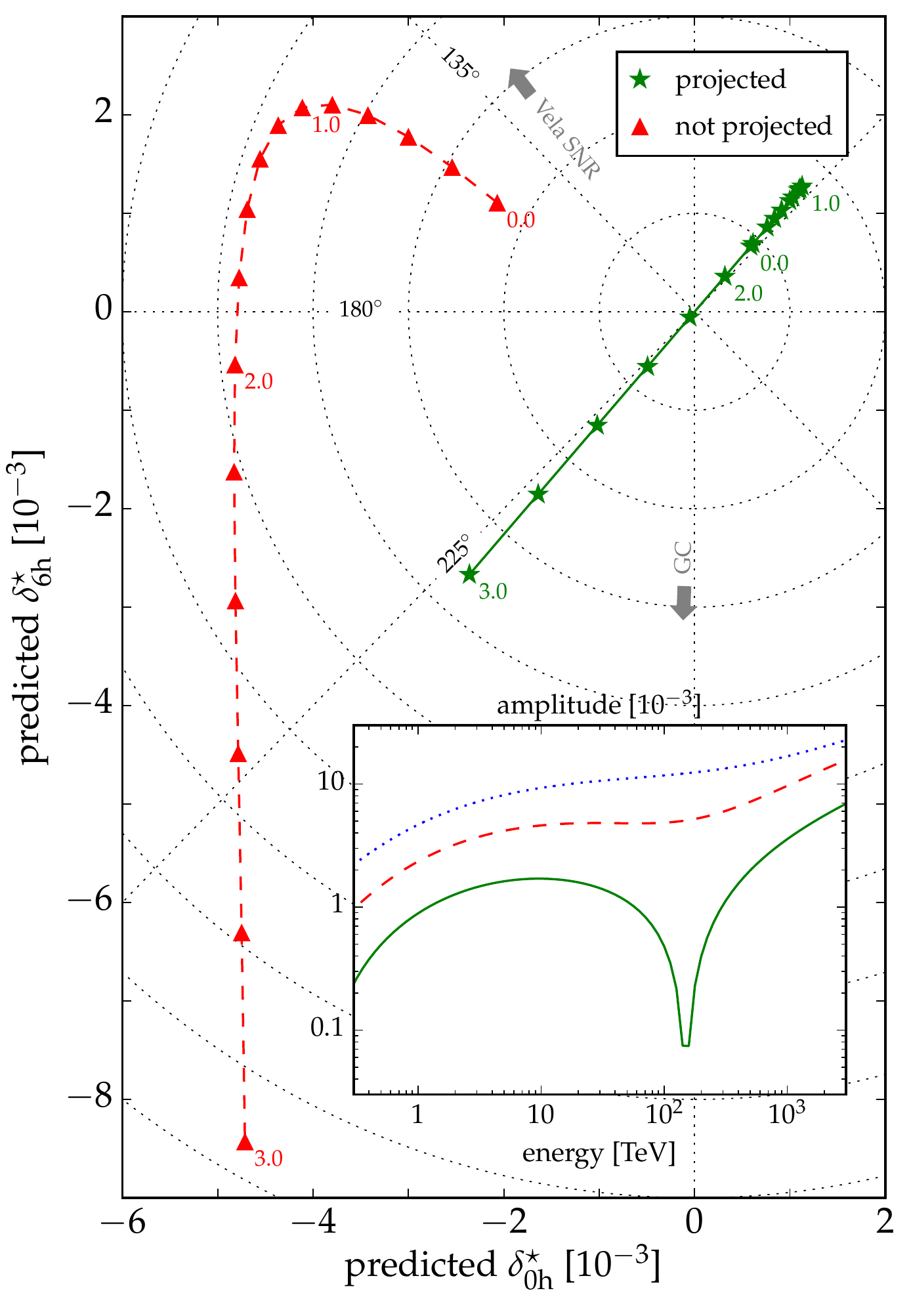}
\caption[]{{\bf Left Panel:} Summary plot of the reconstructed TeV--PeV dipole components $\delta^\star_{0{\rm h}}$ and $\delta^\star_{6{\rm h}}$ in the equatorial plane. The black arrow indicates the Compton-Getting effect from the solar motion with respect to the LSR that we subtracted from the data following Eq.~(\ref{eq:deltarel}). In most cases, the reported dipole amplitude was rescaled to compensate for a bias from the experiment's limited field of view and/or reconstruction techniques. The numbers attached to the data indicate the median energy of the bins as $\log_{10}(E_{\rm med}/{\rm TeV})$. The colored disks show the statistical $1\sigma$ error range. The dashed line and gray-shaded area indicate the magnetic field direction and its uncertainty (projected onto the equatorial plane) inferred from IBEX observations~\cite{Funsten2013}. We also indicate the direction towards the Galactic center (GC). {\bf Right Panel:} Same as the left panel but now showing the predicted dipole anisotropy in a model that accounts for the contribution of local SNRs. The red line shows the prediction assuming isotropic diffusion, where the dipole follows the local CR gradient. The green line shows the same configuration assuming anisotropic diffusion. The inset plot highlights the drop in the dipole amplitude related to the phase-flip around 100~TeV. (Figures from Ref.~\cite{Ahlers:2016njd})}\label{fig:reconstructed}
\end{figure}

Strictly speaking, the prediction (\ref{eq:diffusedipole}) only holds in the rest frame of diffusion (denoted by starred quantities). The relative motion of the solar system with respect to the rest frame of diffusion introduces an apparent dipole anisotropy, which is generally referred to as the Compton-Getting effect~\cite{CG1935}. The dipole in the observer's frame can then be written as the sum
\begin{equation}\label{eq:deltarel}
{\boldsymbol \delta} = {\boldsymbol \delta}^\star + (2+\Gamma){\bf v}/c\,,
\end{equation}
where ${\bf v}$ is the relative velocity of the solar system through the rest frame of diffusion and $\Gamma\simeq2.7$ is the CR spectral index~\cite{Forman1970}. The Compton-Getting effect of the sidereal dipole anisotropy has large uncertainties, related to the ambiguity of identifying the rest frame of CR diffusion. One possible choice is the local standard of rest (LSR) corresponding to our motion towards the solar apex. The velocity vector in Galactic coordinates has been inferred to point towards $l \simeq 47.9^\circ\pm2.9^\circ$ and $b\simeq23.8^\circ\pm2.0^\circ$ with absolute velocity $v_{\rm LSR} \simeq 18.0\pm0.9$ km/s~\cite{Schoenrich2010}. We will use this estimate in the following as a benchmark value. Another choice corresponds to the relative velocity through the local interstellar medium (ISM) towards $l \simeq 5.25^\circ\pm0.24^\circ$ and $b\simeq12.0^\circ\pm0.5^\circ$ with absolute velocity $v_{\rm ISM} \simeq 23.2\pm0.3$ km/s~\cite{McComas2012}. Note that, while the precise value of the sidereal Compton-Getting effect is uncertain, the predicted amplitude is at the level of $3\times10^{-4}$ and almost an order of magnitude smaller than the observed dipole anisotropy at 10~TeV (see Fig.~\ref{fig:dipoledata}).

The field strength of the local ordered magnetic field is expected to be comparable to the turbulent contribution. This implies that the expansion coefficients of the diffusion tensor, Eq.~(\ref{eq:K}), are hierarchical with $\kappa_\perp \ll \kappa_\parallel$ and $\kappa_A \ll \kappa_\parallel$. Therefore, the local diffusion tensor (\ref{eq:K}) reduces to a projector onto the magnetic field direction~\cite{JonesApJ1990,Mertsch:2014cua,Schwadron2014,Ahlers:2016njd}. The local ordered magnetic field on distance scales less than $0.1$~pc can be inferred from the emission of energetic neutral atoms (ENA) from the outer heliosphere observed by the {\it Interstellar Boundary Explorer} (IBEX)~\cite{McComas2009}. The emission of ENA is enhanced along a circular ribbon that defines a magnetic field axis along $l \simeq 210.5^\circ$ and $b\simeq-57.1^\circ$ with an uncertainty of $\sim1.5^\circ$~\cite{Funsten2013}. This agrees with the magnetic field direction inferred from polarization measurements of local stars~\cite{Frisch2015}. 

The left panel of Figure~\ref{fig:reconstructed} shows the reconstructed dipole anisotropy in the equatorial plane (cf.~Eq.~(\ref{eq:ampphase})) inferred from TeV--PeV dipole data. The data are corrected by the predicted Compton-Getting dipole from our relative motion in the LSR. The data in the 1-100~TeV energy range show a strong alignment with the orientation of the local magnetic field projected onto the equatorial plane. This is expected from anisotropic diffusion. Note that the data are only shown with statistical errors. As indicated in the previous section, the limited field of view of observatories is expected to introduce cross-talk between multipole moments that can be responsible for the large scatter between data sets from different observatories. The plot also shows the result of the recent combined analysis of IceCube and HAWC~\cite{TheHAWC:2017uyf} (cf.~Fig.~\ref{fig:IceCubeHAWC}). This measurement is not expected to suffer from strong cross-talk because the combined field of view covers about 95\% of the sky. After correcting for the predicted Compton-Getting shift from our motion in the LSR, the dipole orientation agrees very well with the orientation of the local magnetic field inferred by IBEX.

\begin{figure}[t]\centering
\includegraphics[height=0.52\linewidth]{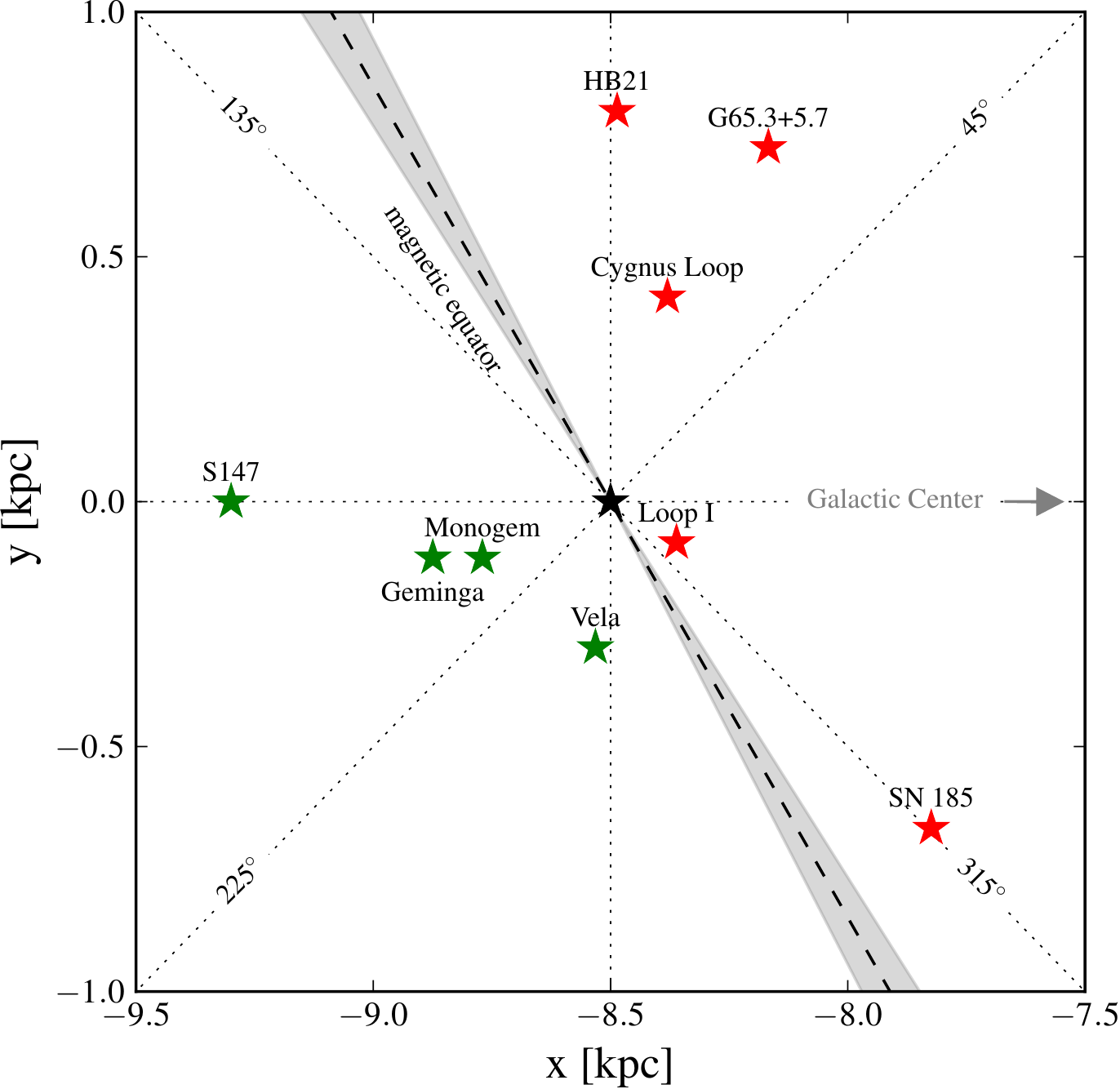}\hfill\includegraphics[height=0.515\linewidth]{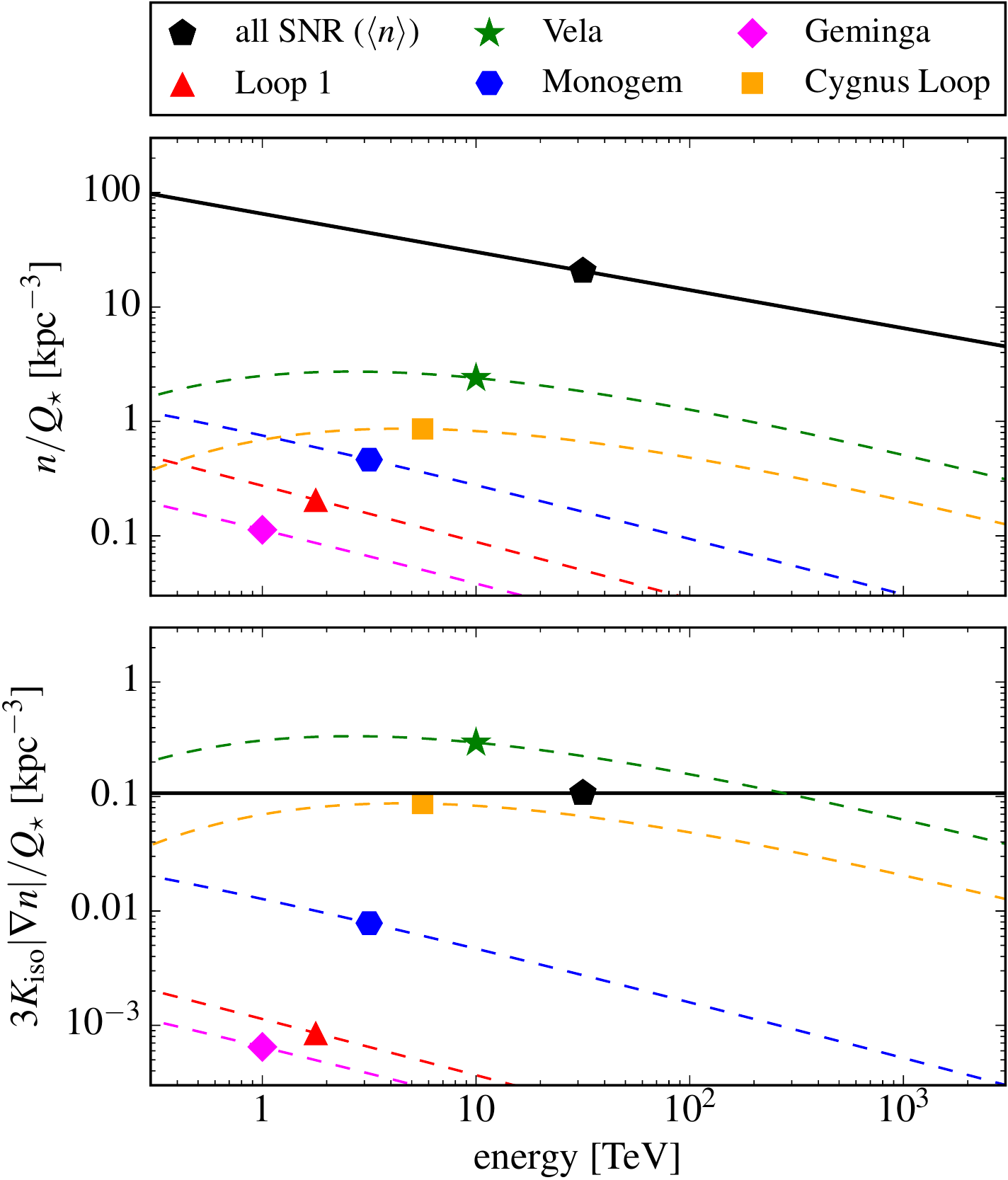}
\caption[]{{\bf Left Panel:} The position of local SNRs in the vicinity of the solar system (black star). The dashed line shows the intersection of the magnetic equatorial plane with the Galactic plane. A CR gradient aligned with the green (red) sources would be visible at a right ascension phase $\alpha\simeq49^\circ$ ($\alpha\simeq229^\circ$) in the model of strong anisotropic diffusion. {\bf Right Panel:} The contribution of the five closest SNRs to the flux (upper panel) and gradient (lower panel) compared to the average contribution of the sum of all Galactic SNRs. (Figure from Ref.~\cite{Ahlers:2016njd})}\label{fig:localSNR}
\end{figure}

The projection of the CR gradient onto the local magnetic field axis leaves only two possible dipole orientations, depending on the relative location of the gradient with respect to the magnetic equator. In the left panel of Fig.~\ref{fig:localSNR} we show the intersection of the magnetic equatorial plane with the Galactic plane together with the location of known nearby SNRs. Assuming that the CR gradient is aligned with the position of one of these potential CR sources, the dipole would be oriented towards $\alpha_1\simeq49^\circ$ for SNRs indicated in green and $\alpha_1\simeq229^\circ$ for SNRs indicated in red. The observed dipole phase below 100~TeV is oriented towards $\alpha_1\simeq49^\circ$, which favors a strong contribution from the local SNRs Vela, Monogem or Geminga. The phase-flip at 100~TeV would indicate the transition of the gradient into the opposite magnetic hemisphere, that also includes the location of the Galactic center.

One can estimate the relative contribution of local SNRs to the CR flux (monopole) and gradient (dipole), taking into account the position and age of these candidate CR sources~\cite{Erlykin:2006ri,Blasi:2011fi,Blasi:2011fm,Pohl:2012xs,Sveshnikova:2013ui,Kumar:2014dma,Savchenko:2015dha}. The calculation of Ref.~\cite{Ahlers:2016njd} assumes impulsive CR emission at the beginning of the Sedov phase (after about $100$~yrs) and CR diffusion with an (effectively) isotropic diffusion coefficient $K_{\rm iso} \simeq 4\times10^{28}(E/3{\rm GeV})^{1/3} {\rm cm}^2/{\rm s}$. The average contribution of all Galactic SNRs assumes a spatial distribution following Ref.~\cite{Case:1998qg} with a source rate of $1/30\,{\rm yr}^{-1}$ and a vertical diffusion height of $3$~kpc. The results of the CR flux and gradient are shown in the right panel of Fig.~\ref{fig:localSNR}. In this model, the flux of local SNRs is subdominant compared to the total contribution of Galactic SNRs, whereas the gradient is dominated below 100~TeV by the Vela SNR. 

The right panel of Fig.~\ref{fig:reconstructed} shows the corresponding dipole projection of this dipole model onto the equatorial plane. The red dashed line shows the predicted dipole under the assumption that the local diffusion tensor is the same as the effective isotropic diffusion tensor assumed for the calculation of the flux and gradient. This is clearly inconsistent with the data shown in the left panel of Fig.~\ref{fig:reconstructed}. However, anisotropic diffusion along the local magnetic field orientation predicts a dipole following the green solid line, in agreement with the behavior seen in the dipole data. The inset plot in the right panel of Fig.~\ref{fig:reconstructed} compares the energy dependence of the amplitude of the full isotropic dipole (blue dotted line), the projected isotropic dipole (red dashed line) and the projected anisotropic dipole (green solid line).

In summary, the observed energy dependence of the dipole amplitude and phase can be well reproduced with diffusion theory after the effect of anisotropic diffusion, local sources, Compton-Getting effect and reconstruction bias are taken into account. A natural candidate for the local source responsible for the phase-flip in the data is the Vela SNR~\cite{Ahlers:2016njd}.

\section{Conclusions}
\medskip

The arrival directions of TeV--PeV Galactic CRs show a rich structure of anisotropies up to a level of one per-mille on various angular scales. In this brief review we have shown that the observed dipole data are consistent with the expectation from diffusion theory, if one accounts for the combined effect of one or more local sources, the presence of a strong ordered magnetic field in our local environment and the reconstruction bias of ground-based observatories. 

The sidereal Compton-Getting effect from our relative motion with respect to the rest frame of diffusion is typically much smaller than the experimental sensitivity if one accounts for systematic uncertainties related to the reconstruction of spherical harmonics coefficients in the limited field of view. However, a recent combined analysis~\cite{TheHAWC:2017uyf} of IceCube and HAWC data with an integrated field of view that covers about 95\% of the sky allows to keep multipole cross-talk under control. The inferred dipole at a median CR energy of 10~TeV aligns well with the local orientation of the magnetic field inferred from IBEX after accounting for our relative motion in the LSR.

The presence of anisotropies at smaller angular scales can be related to the CR transport in the local (turbulent and ordered) magnetic field within one diffusion length. Here, the impact of the heliosphere and the random realization of magnetic turbulence have been invoked to account for hotspots in the anisotropy map (cf.~Fig.~\ref{fig:IceCubeHAWC}) and the excess power down to angular scales of at least $10^\circ$ in the angular power spectrum~\cite{Ahlers:2016rox}. 

\section*{Acknowledgements}
\medskip

I would like to thank the organizers of the {\it 26th Extended European Cosmic Ray Symposium 2018} at Altai State University for their invitation to present this work. I am grateful for many fruitful discussions with and the support of my collaborators Segev Y.~BenZvi, Paolo Desiati, Juan Carlos D{\'i}az-V\'{e}lez, Daniel W.~Fiorino, Philipp Mertsch and the late Stefan Westerhoff. This work was supported by Danmarks Grundforskningsfond (project no.~1041811001) and \textsc{Villum Fonden} (project no.~18994).

\section*{References}

\bibliographystyle{iopart-num}
\bibliography{references}

\end{document}